\documentclass[12pt]{iopart}

\usepackage{citesort}
\usepackage{cite}

\usepackage[breaklinks=true,colorlinks=true,linkcolor=blue,urlcolor=blue,citecolor=blue]{hyperref}
\usepackage{iopams}
\usepackage{graphicx}
\usepackage{subfigure}

	\newcommand{\ket}[1]{\left| #1 \right\rangle}
	\newcommand{\bra}[1]{\left\langle #1 \right|}

\begin{document}
\nocite{*}

\title{Qubit Movement-Assisted Entanglement Swapping}

\author{Sare Golkar$^{1}$}
\address{$^1$Atomic and Molecular Group, Faculty of Physics, Yazd University, Yazd  89195-741, Iran}
\ead{saregolkar@gmail.com}

\author{Mohammad K. Tavassoly$^{1,2}$}
\address{$^1$Atomic and Molecular Group, Faculty of Physics, Yazd University, Yazd  89195-741, Iran
\\ $^2$Photonic Research Group, Engineering Research Center, Yazd University, Yazd  89195-741, Iran}
\ead{mktavassoly@yazd.ac.ir}

\author{Alireza Nourmandipour$^{3}$}
\address{$^3$Department of Physics, Faculty of Science, Sirjan University of Technology, Sirjan, Iran}
\ead{anourmandip@sirjantech.ac.ir}

\begin{abstract}
In this paper, we propose a scheme to generate entanglement between two distant qubits (two-level atom) which are separately trapped in their own (in general) non-Markovian dissipative cavities by utilizing entangling swapping. We consider the case in which the qubits can move along their cavity axes rather than a static state  of motion. We first examine the role of  movement of the qubit by studying the entropy evolution for each subsystem. We calculate the average entropy over the initial states of the qubit. Then by performing a Bell state measurement on the fields leaving the cavities, we swap the entanglement between qubit-field in each cavity into qubit-qubit and field-field subsystems. We use the entangling power to measure the average amount of  swapped entanglement over all possible pure initial states. Our results are presented in two weak and strong coupling regimes.  Our results illustrate the positive role of the movement of the qubits on the swapped entanglement. It is revealed that  by considering certain conditions for the initial state of  qubits, it is possible to achieve a maximally long-leaving stationary entanglement (Bell state) which is entirely independent of the environmental variables as well as the velocity of  qubits. This happens when the two qubits have the same velocities. 
\end{abstract}

\vspace{2pc}

\submitto{J. Phys. B}

\section{Introduction}
In recent decades, there is a great deal of evidence that quantum phenomena play a central role in the development of information theory. Coherent superposition is one of these features which is usually referred as quantum coherence. The non-local quantum correlations among composite subsystems is called entanglement \cite{Horodecki2009QuantumEntanglement}. The importance of quantum entanglement  arises from its various exciting applications such as  quantum teleportation \cite{Braunstein1995}, quantum cryptography \cite{Ekert1991Cryptography}, sensitive measurements \cite{Richter2007},  quantum telecloning \cite{Muarao1999}, superdense coding \cite{Mattle1996} and etc. Due to the rapid growth of the applications of these kinds of quantum states in quantum information processing implementations, a great deal of attention has been devoted to the generation and detection of entangled states. Most of these proposals rely on the interaction of atoms (real or artificial) with optical cavities \cite{Raimond2001}. Other proposals include quantum dots \cite{Pazy2003}, atomic ensembles \cite{Julsgaard2001}, superconducting quantum interference devices \cite{Yang2003,Yang2004}, photon pairs \cite{Aspect1981}, superconducting qubits \cite{Izmalkov2004} and trapped ions \cite{Pachos2002,Cirac1995,Sackett2000experimental}.

Approximately, all of the introduced schemes depends on the interactions (direct or indirect) between subsystems. For instance, it has been shown that the Jaynes-Cummings model (JCM)  which describes the interaction of atoms (two- or multi-level) with cavity field \cite{Jaynes1963}, could generate entanglement between an atom and a quantized field. This model has been extended to include the interaction of multiple atoms with a  multi-mode electromagnetic field  \cite{Tavis1968}. Thanks to the nonlocality of quantum correlations, it is possible to entangle two or more particles which are distributed over distances without any interactions and, or common history. This phenomenon is called entanglement swapping  \cite{Zukowski1993}. In this protocol, the basic recipe is to make a more general system. Then by projection of  the quantum state of the whole system onto a maximally entangled Bell state, it is possible to swap the entanglement between these subsystems. There have been many works on this interesting topic. For instance, in \cite{Polkinghorne1999} it has been generated for continuous-variable systems.  Multiparticle entanglement swapping has been studied in \cite{Bose1998}. The unconditional entanglement swapping has been experimentally demonstrated in \cite{Jia2004}. In \cite{Hu2011}, this phenomenon has been discussed by using a quantum-dot spin system. It also has been shown that entanglement swapping could be used for the optimization of entanglement purification \cite{Shi2000}. One-cavity scheme enabling to implement delayed choice for entanglement swapping in cavity QED has been investigated in \cite{Almeida2015}. The effect of detuning and Kerr medium on the entanglement swapping has been studied in \cite{Ghasemi2017}. Recently, a high-fidelity, unconditional entanglement swapping experiment in a superconducting circuit has been performed in \cite{Ning2019}. Very recently, the swapping of entangled states between two pairs of photons emitted by a single quantum dot has been performed experimentally \cite{zopf2019entanglement}. 

On the other hand, contrary to the closed systems which are ideal, the real physical systems are open. This means that dissipation is always present in those systems. Actually, the inescapable interaction between the aim system and its surrounding environment makes  the entanglement fragile \cite{Yu2003,Yu2004}. Because a long-lasting entangled state is an essential resource for the quantum information theory, many strategies have been devoted to fight against the destructive environmental effects: the theory of open quantum systems
\cite{Nourmandipour2015b,Mortezapour2011,kim2012protecting,Rafiee2016,Rafiee2017,Mortezapour201826,mortezapour2018protecting}. However, it should be noted that the idea of  interaction of a quantum system with the surrounding environment is not always bad. For instance, it has been shown that there exists a long-living entangled state due to the interaction of a two-qubit system with a common environment \cite{Nourmandipour2015,Maniscalco2008}. This idea has been generalized to an arbitrary number of qubits inside an environment \cite{Nourmandipour2016,Nourmandipour2016J,Shankar2014}. Moreover, quantum reservoir engineering  has been proven to be useful in stabilizing open quantum systems \cite{Schirmer2010} and remote entanglement and concentration \cite{Didier2018}, etc.  Recently, it has been shown that an external classical field is a practical scheme to preserve the entanglement between two dissipative systems \cite{Xiao2010}. In this regard, many studies such as non-Markovianity \cite{Haikka2010}, quantum speedup \cite{Zhang2015pra} quantum Fisher information  \cite{Ren2016}, etc. have been studied.   

Recent experimental schemes in quantum information processing rely on the control of single qubits inside (optical) cavities. However, in practical implementations, achieving a static state of qubits in a cavity is a difficult task, if not impossible! In a pioneering work, the effect of the movement of two qubits inside non-Markovian environments on the protection of the initial entanglement has been studied in \cite{Mortezapour2017}. Moreover, other studies illustrate the effect of movement of qubits (both uniform and accelerated) on the interaction between such qubits and electromagnetic radiation \cite{Calaj2017,garcia2017entanglement}. This includes the relativistic velocities for qubits \cite{Felicetti2015,Moustos2017}. In a very recent paper, it has been shown the positive role of  movement of  qubits on the entanglement dynamics of an arbitrary number of qubits in a Markovian and/or non-Markovian environment \cite{golkar2019entanglement}. It also has been shown that when all of the qubits have the same velocity, the stationary state of  entanglement is independent of the velocity of  qubits \cite{golkar2019entanglement}. 

All of the statements mentioned above motivate us to examine the effect of  movement of qubit on the entanglement swapping between two separate subsystems. To end this, we consider two independent cavities each contains a moving two-level system (qubit) in the presence of dissipation. We model the environment as a  set of infinite quantized harmonic oscillators. We take the situation in which each qubit is allowed to move along the cavity axis. We also consider the non-relativistic velocities for qubits. In this situation, the exact dynamics for each subsystem is obtained for both weak and strong coupling regimes corresponding to bad and good cavity limits. We first explore the role of  movement of the qubit on the entropy evolution for each subsystem. After that, we perform a Bell state measurement (BSM) on the cavity fields leaving the cavities. This swaps the entanglement between the qubit and the field in each cavity into qubit-qubit and field-field entanglements. We use the concurrence measure \cite{Wootters1998} to quantify the amount of swapped entanglement. Naturally, this depends on the initial state of the qubits. Our parametrization for the initial states of qubits allows us to establish an input-independent dynamics of entanglement by taking a statistical average over the initial states of two qubits. This is called entangling power which originally introduced for unitary maps \cite{Zanardi2000} and then  generalized for dissipative channels \cite{Nourmandipour2016QIC}.

The rest of the paper is organized as follow. In Sec. \ref{SecModel}, the model and the various related parameters are introduced and the exact dynamics of each subsystem is obtained. Section \ref{secAtFiEnt} deals with the effect of the movement of the qubits on the entropy evolution of each subsystem. We study the entanglement swapping phenomena in details in Sec. \ref{SecEntSwa}. Finally, in Sec. \ref{Sec.Con} we summarize the paper.

\section{Exact dynamics of the single moving qubit system in dissipative regime}
\label{SecModel}

\begin{figure}[ht]
   \centering
\includegraphics[width=0.5\textwidth]{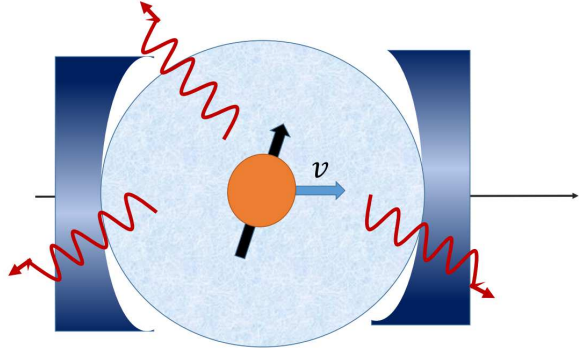}
   \caption{\label{Fig1} Pictorial representation of a setup in which a moving qubit is interacting with a dissipative cavity.}
  \end{figure}

We consider two similar, but separate dissipative cavities, each containing a moving two-level atom (qubit) with  an excited (ground) $\ket{e}$ ($\ket{g}$) state. These states are separated by transition frequency $\omega_{\text{qb}}$.  Each qubit is taken to move along the z-axis of the corresponding cavity with constant velocity $v$ (see Fig. \ref{Fig1}). The movement of  qubits is characterized by a sine term due to the boundary conditions.
 The Hamiltonian for each system (with $\hbar = 1$) is given by
 \begin{eqnarray}
 \label{eq:model}
   \hat H_{\text{(AF)}_i}&=&\hat{H}_{{\text{(AF)}}_i}^0+ \hat{H}_{{\text{(AF)}}_i}^{\text{Int}},\;\;\;\;\;\;\;i=1,2\;\;
  \end{eqnarray}
in which
 \begin{subequations}
     \begin{eqnarray} 
   \hat{H}_{{\text{(AF)}}_i}^0&=&\sum_{k} \omega_{k} \hat{a}^{\dagger}_{k} \hat{a}_{k}  + \frac{\omega_{\text{qb}}}{2}\hat{\sigma}_{z} ,\\
  \hat{H}_{{\text{(AF)}}_i}^{\text{Int}}&=&\sum_{k} \alpha_ig_{{k}} f_{k}(z_{i})\hat{a}_{k}\hat{\sigma}^{+}+\text{H.c.}
    \end{eqnarray}
 \end{subequations}
where ${{\hat{\sigma }}_{z}}=\left| e \right\rangle \left\langle  e \right|-\left| g \right\rangle \left\langle  g \right|$ is the Pauli matrix, ${{\omega }_{k}}$  denotes the frequency of the cavity quantized modes, while ${{\hat{a}}_{k}}$ ($\hat{a}_{k}^{\dagger}$) are the annihilation (creation) operators of the cavity $k$th mode. ${{g}_{k}}$ denotes the coupling constant between the qubit and the $k$th mode of  environment. The interaction of the $i$th qubit with the environment is measured by the dimensionless constant $\alpha_i$. ${{\hat{\sigma }}_{+}}=\left| e \right\rangle \left\langle  g \right|$ (${{\hat{\sigma }}_{-}}=\left| g \right\rangle \left\langle  e \right|$) denotes the qubit raising (lowering) operator. Furthermore, the function  $f_k(z_{i})$ describes the shape function of the $i$th qubit motion along z-axis \cite{PhysRevA.70.013414}. In this regard, this parameter is given by \cite{katsuki2013all}
  \begin{equation}
  f_{k}(z_{i})=f_{k}(v_{i}t)=\sin[\omega_{k}(\beta_it-\Gamma)], \ \ \ \  i=1,2
  \label{eq:velshapediss}
  \end{equation}
 where, $\beta_i=v_i/c$ and $\Gamma=L/c$ with $L$ being the size of  cavity and $c$ is the velocity of light. The sine term in the above relation comes from the boundary conditions. It should be noted that the translational motion of the qubits has been treated classically $(z=vt)$. This is the situation in which the de Broglie wavelength $\lambda_B$ of the qubits is much smaller than the wavelength $\lambda_0$ of the resonant transition (i.e., $\lambda_B/\lambda_0\ll 1$) \cite{Mortezapour2017} which means that $\beta_i\ll 1$.

Formally, it is more convenient to work in the interaction picture. The Hamiltonian (\ref{eq:model}), in the interaction picture is given by
\begin{equation}\label{int}
 \hat{{\mathfrak V}}_{\text{(AF)}_i}= e^{i \hat{H}_{\text{(AF)}_i}^0t }\; \hat{H}^{\text{Int}}_{\text{(AF)}_i}\;e^{-i \hat{H}_{\text{(AF)}_i}^0t }.
  \end{equation}
    After some manipulation, the explicit form of the Hamiltonian in the interaction picture may be obtained as  
      \begin{equation}\label{intpicdiss}
       \hat{{\mathfrak V}}_{\text{(AF)}_i}=\sum_{k}\alpha_ig_{{k}} f_{k}(z_{i})\hat{a}_{k}\hat{\sigma}^{+} e^{-i(\omega_{k}-\omega_{\text{qb}})t}+\text{H.c.}
        \end{equation} 
In the above relation, without loss of generality,we consider the case in which  each qubit has the same speed, i.e., $v_1=v_2\equiv v$ ($\beta_1=\beta_2\equiv\beta$) and also the constants $\alpha_1$ and $\alpha_2$ be the same, i.e., $\alpha_1=\alpha_2\equiv\alpha$.

We suppose that there is no excitation in the cavities before the occurrence of interaction and each atom is in the coherent superposition of the exited $\ket{e}$ and  ground state $\ket{g}$ as
\begin{equation}
\ket{\psi(0)}_i=\left( \cos(\theta_i/2) \ket{e}+\sin(\theta_i/2) e^{i\phi_i}\ket{g}\right)\ket{\boldsymbol{0}}_{R},
\label{eq:initialstatediss}
\end{equation}
in which $\ket{\boldsymbol{0}}_{R}$ is the multi-mode vacuum state of the cavity.
 In the above relation $\theta_i\in\left[ 0,\pi\right] $ and $\phi_i\in\left[ 0,2\pi\right] $ for $i=1,2$.
According to Hamiltonian  (\ref{intpicdiss}), the quantum state of the $i$th system at any time $t$ can be written as 
\begin{equation}
\begin{aligned}
\ket{\psi(t)}_i&=\cos(\theta_i/2){\cal E}(t)\ket{e}\ket{\boldsymbol{0}}_{R}+\sin(\theta_i/2) e^{i\phi_i}\ket{g}\ket{\boldsymbol{0}}_{R}\\
&+\cos(\theta_i/2)\sum_{k} {\cal G}_{k}(t)\ket{g}\ket{\boldsymbol{1_{k}}}
\end{aligned}
\label{eq:statediss}
\end{equation}
where  $\ket{\boldsymbol{1_{k}}}$ describes the state of the environment with only one photon in the $k $th mode.

Using the time-dependent Schr\"{o}dinger equation in the interaction picture,we are readily led to integro-differential equation for the amplitude ${\cal E}(t)$
  \begin{equation}
 \dot{\cal E}(t)= -\alpha^2\int_{0}^{t}\! F(t,t'){\cal E}(t')  \, \mathrm{d}t',
 \label{eq:survxidiss}
  \end{equation}
where the kernel $F(t,t')$ takes the form
\begin{eqnarray}\label{Correlation functiondiss}
F(t,t')=\sum_{k}|g_k|^2e^{i\delta_k(t-t')}f_k(vt)f_k(vt'),
\end{eqnarray}
in which, $\delta_k=\omega_{\text{qb}}-\omega_k$. 

As is seen, ${\cal E}(t)$ depends on the spectral density as well as the shape function of the qubit motion. In the continuum limit for the reservoir spectrum, the
sum over the modes is replaced by the integral
\begin{equation}
\sum_{k}|g_k|^2\rightarrow \int \mathrm{d}\omega J(\omega)
\end{equation}
where $J(\omega)$ is the reservoir spectral density.
 The nonperfect reflectivity of the cavity mirrors implies a Lorentzian spectral density for the cavity as follows \cite{Maniscalco2008}
\begin{equation}
\label{eq:specden}
J(\omega)=\frac{W^2}{\pi}\frac{\lambda}{(\omega-\omega_{\text{qb}})^2+\lambda^{2}},
\end{equation}
where we have assumed that the qubit interacts resonantly with the center frequency of the
cavity modes.  The weight $W$ is proportional to the vacuum Rabi frequency and $\lambda$ is the width of the distribution which describes the cavity losses. 

Using \eqref{eq:specden} in \eqref{Correlation functiondiss}, we arrive at the following expression for $F(t,t')$
  \begin{equation}
  F(t,t')=\frac{W^2\lambda}{\pi}\int\text{d}\omega\frac{\sin[\omega(\beta t-\Gamma)]\sin[\omega(\beta t'-\Gamma)]}{(\omega-\omega_{\text{qb}})^2+\lambda^{2}}e^{-i(\omega-\omega_0)(t-t')}.
  \end{equation}
  Again, in the continuum limit (i.e., $\Gamma\rightarrow\infty$) \cite{park2017protection}, the analytical solution of the above relation gives rise to
  \begin{equation}
  F(t,t')=\frac{W^2}{2}e^{-\lambda(t-t')}\cosh[\beta\bar{\lambda}(t-t')],
  \end{equation}
  in which $\bar{\lambda}\equiv\lambda+i\omega_{\text{qb}}$. Once again $F(t,t')=G(t-t')$, which motivates us to use the Laplace transformation technique. After some straightforward, but long manipulations, we may obtain the analytical solution of \eqref{eq:survxidiss} as follows
  \begin{equation}
  \label{eq:suramplitdiss}
  \begin{aligned}
{\cal E}(t)&=\frac{(q_1+y_+)(q_1+y_-)}{(q_1-q_2)(q_1-q_3)}e^{q_1\lambda t} \\
&+\frac{(q_2+y_+)(q_2+y_-)}{(q_2-q_1)(q_2-q_3)}e^{q_2\lambda t} \\
&+\frac{(q_3+y_+)(q_3+y_-)}{(q_3-q_1)(q_3-q_2)}e^{q_3\lambda t}.
  \end{aligned}
  \end{equation}
  in which the quantities $q_i \ \ (i=1,2,3)$ are now the solutions of the cubic equation
\begin{equation}
\label{eq:cubicdiss}
q^3+2q^2+(y_+y_-+\frac{R^2}{2})q+\frac{R^2}{2}=0
\end{equation}
with $y_{\pm}=1\pm\beta(1+i\omega_{\text{qb}}/\lambda)$ and $R={\cal R}/\lambda$  with vacuum Rabi frequency ${\cal R}=\alpha W$. 

\section{Atom-Field Entanglement of Subsystems}
\label{secAtFiEnt}

In the previous section, we have solved the SchrÃ¶dinger equation for the case of a moving qubit inside a cavity in both non-dissipative and dissipative regimes.  Before we consider the entanglement swapping phenomena, we intend to illustrate the effect of the movement of the qubits on the entanglement dynamics. Among the various measures for computing the degree of entanglement between bipartite systems, we use the linear entropy which is defined as \cite{Peters2004}
\begin{equation}\label{linear}
S_{A}(\theta,\phi;t)=1-\text{Tr}\left( \hat{\rho}_{_A}^2\right) ,
\end{equation}
in which $\hat{\rho}_{_A}$ is the atomic reduced density matrix for each subsystem. We notice that we have omitted the subscript $i$ (with $i=1,2$) from parameters $\theta$ and $\phi$, because we are dealing with only one subsystem.  The linear entropy can range between zero, corresponding to a completely pure state, and $(1-1/d)$ corresponding to a completely mixed state, in which $d$ is the dimension of the density matrix (here, $d=2$).

Using Eqs. (\ref{eq:statediss}) and (\ref{linear}), the linear entropy  at any time can be derived as:
\begin{equation}\label{Lin}
S_{A}(\theta,\phi;t)=2\left(1-\left|{\cal E}(t)\right| ^2  \right) \left|{\cal E}(t)\right| ^2\cos^4(\theta/2)
\end{equation}
which does not depend on parameter $\phi$. It is evident that the maximum amount of linear entropy is obtained for $\theta=2m\pi$ with $m=0,1,2,\dots$ which corresponds to the situation in which the qubit is initially in the excited state. Also, one observes that the linear entropy is zero for the qubit  which is initially in the ground state (i.e., $\theta=(2m+1)\pi$ with $m=0,1,2,\dots$). For other values of $\theta$, the behaviour of the linear entropy is the same but with different amplitude. This leads us to the natural question: on average, how much linear entropy is obtained over all initial states. This provides an input-independent dynamics.  This can be done by computing the average of linear entropy with respect to all possible input states on the surface of the Bloch sphere as:
\begin{equation}
S_{_A}^{\text{av}}(t)=\int S_A(\theta,\phi;t) \, \text{d}\Omega,
\end{equation}
in which $\text{d}\Omega$ is the normalized $\text{SU}(2)$ Haar measure,
\begin{equation}\label{Omega}
\text{d}\Omega=\frac{1}{4\pi}\sin\theta\text{d}\theta\text{d}\phi.
\end{equation}
From Eqs. \eqref{Lin}-\eqref{Omega}, it can be easily shown that
\begin{equation}\label{AveLin}
S_{_A}^{\text{av}}(t)=\frac{2}{3}\left(1-\left|{\cal E}(t)\right| ^2  \right) \left|{\cal E}(t)\right| ^2
\end{equation}

\begin{figure}[h!]
\centering
\subfigure[\label{Fig2a} Linear entropy for $\theta=0$ ]{\includegraphics[width=0.4\textwidth]{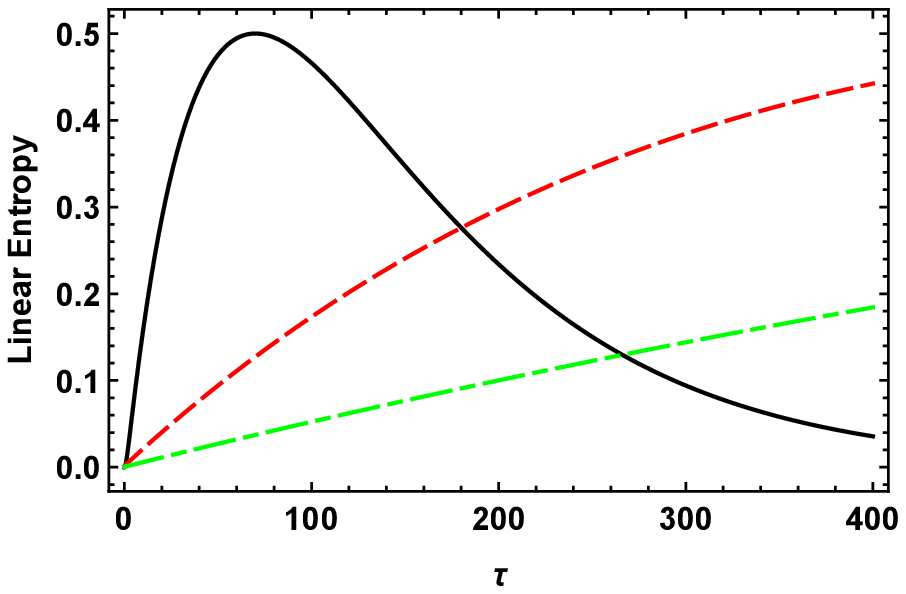}}
\hspace{0.05\textwidth}
\subfigure[\label{Fig2b} Average of linear entropy ]{\includegraphics[width=0.4\textwidth]{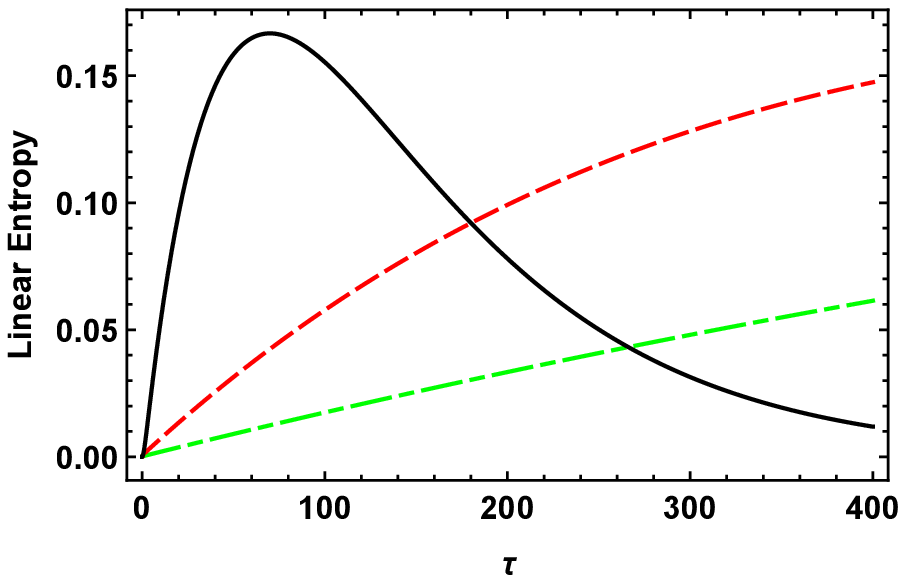}}

\caption{Time evolution of (a) linear entropy for $\theta=0$ and (b) the average of linear entropy as functions of scaled time $\tau$ for weak coupling regime, i.e., $R=0.1$ with $\beta=0$ (solid black line),  $\beta=2\times 10^{-9}$ (dashed red line) and $\beta=4\times 10^{-9}$ (dot-dashed green line). In these plots we have set $\omega_0/\lambda=1.5\times 10^{9} $.} \label{Fig2}
   \end{figure}

Fig. \ref{Fig2} illustrates the linear entropy \eqref{Lin} for $\theta=0$ and the average of linear entropy \eqref{AveLin} in the weak coupling regime ($R=0.1$) for different values of velocity of the qubit. Taking a glance at these plots reveals the positive role of the movement of the qubit on the survival of the linear entropy. As is seen, by increasing the velocity of the qubit, the linear entropy reaches its maximum value at longer times. The decaying behaviour of linear entropy represents a Markovian process. The maximum amount of linear entropy is obtained for $\theta=0$. For $\beta=0$, we recover the results presented in \cite{NourmandipourPRA2016}.  

\begin{figure}[h!]
\centering
\subfigure[\label{Fig3a} Linear entropy for $\theta=0$ ]{\includegraphics[width=0.4\textwidth]{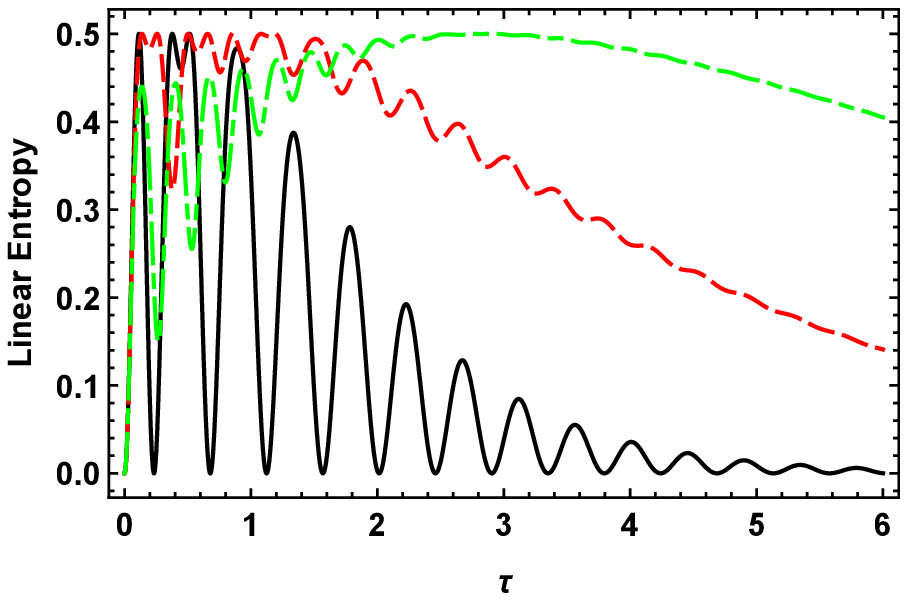}}
\hspace{0.05\textwidth}
\subfigure[\label{Fig3b} Average of linear entropy]{\includegraphics[width=0.4\textwidth]{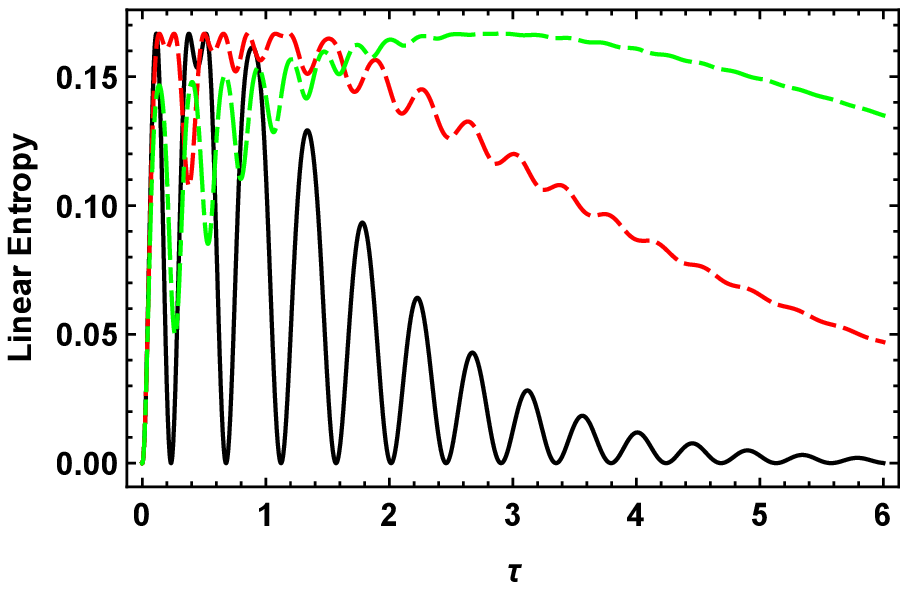}}

\caption{Time evolution of (a) linear entropy for $\theta=0$ and (b) the average of linear entropy as functions of scaled time $\tau$ for strong coupling regime, i.e., $R=10$ with $\beta=0$ (solid black line),  $\beta=10\times 10^{-9}$ (dashed red line) and $\beta=15\times 10^{-9}$ (dot-dashed green line). Again $\omega_0/\lambda=1.5\times 10^{9} $.} \label{Fig3}
   \end{figure}

In Fig. \ref{Fig3} we have plotted the linear entropy for $\theta=0$ and the average of linear entropy in the strong coupling regime ($R=10$) for several motion situations of the qubit. Again, the positive role of the movement of  qubit on the survival of the linear entropy is quite clear. Increasing the velocity of the qubit not only makes the linear entropy survives at longer times but also washes out the oscillatory behaviour of the linear entropy. These oscillations are a sign of the non-Markovian process. Actually, in this coupling regime, the interaction between the qubit and its surrounding environment  is so strong that part of the information that has been taken by the environment is fed back to the qubit. As is clear, in the presence of the movement of the qubit, the sudden death of linear entropy is no longer seen. Again, $\beta=0$ recovers the results in \cite{NourmandipourPRA2016}. 

\section{Entanglement Swapping}
\label{SecEntSwa}

\begin{figure}[ht]
   \centering
\includegraphics[width=0.6\textwidth]{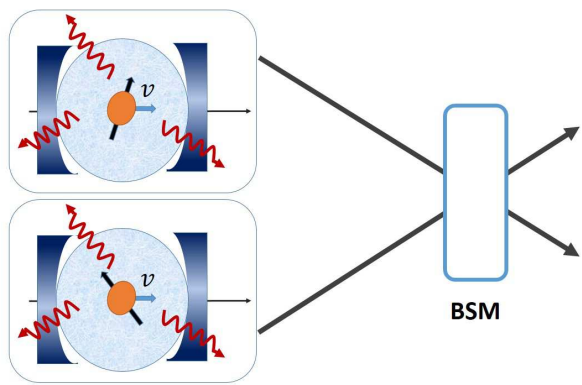}
   \caption{\label{Fig4} Pictorial illustration of the entanglement swapping phenomenon. In each subsystem, the moving qubit is interacting with its own environment. Because of the nonperfect reflecting mirrors of the cavities, the photons can leak out of the cavities. Then a Bell state measurement is performed on these photons establishes the entanglement between the qubits.
}
  \end{figure}

 In the previous section, we examined the positive role of the movement of a single qubit on the entropy evolution of the qubit and its surrounding environment. In this section, we consider two similar but separable systems introduced in Sec. \ref{SecModel}. The time evolution of each system is given in Eq. \eqref{eq:statediss}. As one expects, their states are separable, i.e.,
         \begin{equation}\label{statet}
         \hat{\rho}(t)= \ket{\Psi(t)}\bra{\Psi(t)},
       \end{equation}
 in which $\ket{\Psi(t)}=\ket{\psi(t)}_1\otimes\ket{\psi(t)}_2$. However, the qubit-qubit entangled states are more important due to their applications in quantum information processing. Therefore, it is quite logical to search a strategy to exchange the entanglement stored between qubit-field in each system into qubit-qubit and field-filed entanglement, as illustrated in Fig. \ref{Fig4}. This can be done by projection $\ket{\Psi(t)}$ onto one of the Bell states of the cavity fields. According to the wave function \eqref{eq:statediss}, one observes that there exist only the vacuum and first excited states of the field modes in each system. Therefore, it is possible to consider the following Bell-like states of the fields \cite{Lee2013}: 
 \begin{subequations}
 \begin{eqnarray}
  \ket{\psi^{\pm}}_\mathrm{F}&=&\frac{1}{\sqrt{2}}\left( \ket{\boldsymbol{0}}_{R_1}\ket{\boldsymbol{1}}_{R_2}\pm\ket{\boldsymbol{1}}_{R_1}\ket{\boldsymbol{0}}_{R_2} \right), \\
  \ket{\phi^{\pm}}_\mathrm{F}&=&\frac{1}{\sqrt{2}}\left( \ket{\boldsymbol{0}}_{R_1}\ket{\boldsymbol{0}}_{R_2}\pm\ket{\boldsymbol{1}}_{R_1}\ket{\boldsymbol{1}}_{R_2} \right),
 \end{eqnarray}
 \end{subequations}
 in which $\ket{\boldsymbol{0}}_{R_i}$ has been defined before and 
    \begin{equation}
    \ket{\boldsymbol{1}}_{R_i}\equiv\sum_{k}\Theta_k\ket{\boldsymbol{1}_{k}}_i
    \end{equation}
    where $\sum_k |\Theta_k|^2 =1$ with $\Theta_k$ is related to the pulse shape associated with the incoming photons. The next step is to construct the projection operator  $P_\mathrm{F}=\ket{M}{}_\mathrm{FF}\bra{{}M}$ in which $M\in\left\lbrace \psi^{\pm},\phi^{\pm}\right\rbrace$. Consequently, operating one of this projection operators onto $\ket{\Psi(t)}$ leaves the field states in the corresponding Bell-type state and also establishes entangled atom-atom state.

In the continuation, we consider the following projection operator
\begin{equation}
P^-_\mathrm{F}=\ket{\Psi^-}{}_\mathrm{FF}\bra{{}\Psi^-}.
\end{equation}
Then we have
\begin{equation}
P^-_F \ket{\Psi(t)}=\ket{\Psi^-}{}_\mathrm{F}\otimes\ket{\Psi_\mathrm{AA}(t)}
\end{equation}
in which $\ket{\Psi_\mathrm{AA}(t)}$ is the qubit-qubit state (after normalization):
\begin{equation}\label{AAunnormPsi}
\begin{aligned}
     \ket{\Psi_\mathrm{AA}(t)}& =\frac{1}{\sqrt{{\cal N}(t)}}\left\lbrace  X(\theta_1,\theta_2,t)\Big(\ket{e}\ket{g}-\ket{g}\ket{e}\Big)\right. \\
     &+\left.  Y(\theta_1,\theta_2,\phi_1,\phi_2) \ket{g}\ket{g}\right\rbrace,
\end{aligned}
\end{equation}
where the normalization coefficient reads as
\begin{equation}\label{NorCoeffPsi}
\begin{aligned}
     {\cal N}(t)&= 2|X(\theta_1,\theta_2,t)|^2+|Y(\theta_1,\theta_2,\phi_1,\phi_2)|^2.
\end{aligned}
\end{equation}
In the above relations, we have defined
\begin{subequations}
\begin{eqnarray}
X(\theta_1,\theta_2,t)&=&\cos(\theta_1/2)\cos(\theta_2/2) {\cal E}(t), \label{eq:X} \\
Y(\theta_1,\theta_2,\phi_1,\phi_2)&=&\sin(\theta_1/2)\cos(\theta_2/2)e^{i\phi_1} \nonumber \\  &-&\sin(\theta_2/2)\cos(\theta_1/2)e^{i\phi_2}. \label{eq:Y}
\end{eqnarray}
\end{subequations}
In what follows, we use concurrence as the figure of merit for the amount of entanglement between the two qubits. It has been defined as \cite{Wootters1998}
       \begin{equation}\label{con}
        E\left( \hat{\rho}(t)\right) =\mathrm{max}\{0,\sqrt{\lambda_1}-\sqrt{\lambda_2}-\sqrt{\lambda_3}-\sqrt{\lambda_4}\},
       \end{equation}
where $\lambda_i$, $i=1,2,3,4$ are the eigenvalues (in decreasing order) of the  matrix
$\hat{\rho}_{_\mathrm{AA}}\left(\sigma_1^y\otimes\sigma_2^y\hat{\rho}_{_\mathrm{AA}}^{*}\sigma_1^y\otimes\sigma_2^y\right)$ with $\hat{\rho}_{_\mathrm{AA}}^*$ the complex conjugate of $\hat{\rho}_{_\mathrm{AA}}$ and $\sigma_k^y:=i(\sigma_k-\sigma_k^\dag)$.  The concurrence varies between 0 (completely separable) and 1 (maximally entangled).
For the state (\ref{AAunnormPsi}),  this parameter reads as
\begin{equation}
E\left( \hat{\rho}(t)\right)=\dfrac{2|X(\theta_1,\theta_2,t)|^2}{2|X(\theta_1,\theta_2,t)|^2+|Y(\theta_1,\theta_2,\phi_1,\phi_2)|^2}.
\label{conexPsi}
\end{equation}
A glance at the resulting concurrence reveals that the concurrence \eqref{conexPsi} does not depend on the shape of the incoming photons.  Also, it  is evident that whenever $|Y(\theta_1,\theta_2,\phi_1,\phi_2)|^2=0$, the concurrence would be independent of time and it always remains  at its maximum value, i.e., 1. According to Eq. (\ref{eq:Y}), it amounts to solve the following relation,
\begin{equation}
\dfrac{1}{2}\Big( 1-\cos\theta_1\cos\theta_2 -\sin\theta_1\sin\theta_2\cos(\phi_1-\phi_2)\Big)=0.
\end{equation}
The above relation is fulfilled with the following set of solutions
\begin{itemize}
\item $\theta_1 =\theta_2=2n\pi$ and arbitrary values of $\phi_1$ and $\phi_2$ with $n=0,1,2,\cdots$
\item $\theta_1 =\theta_2 \ \text{and} \ \phi_1-\phi_2=2m\pi, \  \text{with} \ m=0,\pm 1$.
\end{itemize}
These conditions lead to the maximally entangled Bell state (up to an irrelevant global phase)
  \begin{eqnarray}\label{Bellstate1}
  \ket{\Psi^-}=\frac{1}{\sqrt{2}}(\ket{e}\ket{g}-\ket{g}\ket{e}).
  \end{eqnarray}
On the other hand, for  $\theta_1 \ \text{or} \ \theta_2=(2n+1)\pi$ and arbitrary values of $\phi_1$ and $\phi_2$, the concurrence is always zero and we have the qubit-qubit state $\ket{g}\ket{g}$ as the stationary state. 
   
Again, similar to the previous section, we can establish an input-independent dynamics for the swapped entanglement, which is called entangling power. This is done by taking a statistical average over all initial states \cite{Nourmandipour2016QIC}
   \begin{equation}
   {\mathfrak E}(t):=\int E\left(\rho(t)\right) \, d\mu( |\psi(0)\rangle),
   \label{eq:enpower}
   \end{equation}
   where $ d\mu( |\psi(0)\rangle)$ is the probability measure over the submanifold of product states in $\mathbb{C}^2\otimes \mathbb{C}^2$. The latter is induced by the Haar measure of ${\rm SU}(2) \otimes {\rm SU}(2)$. Specifically, referring to the parametrization of (\ref{eq:initialstatediss}), it reads
   \begin{equation}
   d\mu( |\psi(0)\rangle)=\frac{1}{16\pi^2}\prod\limits_{k=1}^2 \sin\theta_k\text{d}\theta_k\text{d}\phi_k.
   \end{equation}
   According to the above definition, the entangling power $\mathfrak E$ is normalized to 1. It is trivial that in this case, it lies in $[0,1]$.\\

\begin{figure}[h!]
   \centering
\includegraphics[width=0.4\textwidth]{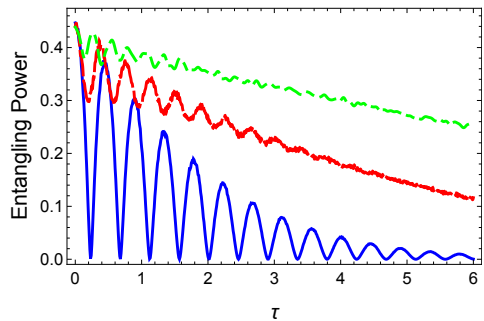}

\caption{Time evolution of the entangling power of the atom-atom state after BSM ($P^-_\mathrm{F}=\ket{\Psi^-}{}_\mathrm{FF}\bra{{}\Psi^-}$) as a function of scaled time $\tau$ for strong coupling regime, i.e., $R=10$ with $\beta=0$ (solid black line),  $\beta=10\times 10^{-9}$ (dashed red line) and $\beta=15\times 10^{-9}$ (dot-dashed green line). We have set $\omega_0/\lambda=1.5\times 10^{9} $.} \label{Fig5}
   \end{figure}
   
Figure \ref{Fig5} illustrates the entangling power as a function of scaled time $\tau=\lambda t$ in the strong coupling regime (i.e., $R=10$) for different values of $\beta$. For $\beta=0$, we recover the results presented in \cite{NourmandipourPRA2016}. In this case, entangling power has an oscillating behaviour which is a characteristic feature of non-Markovian regime. Entanglement sudden death phenomenon is clearly seen.  As is observed, the entangling power exhibits an oscillatory decay behaviour in the absence and presence of the movement of the qubits. However, the presence of movement of qubits makes the decay of  entanglement becomes slow. This can be understood by paying attention  to the fact that the entanglement between the two qubits depends directly on the entanglement between the qubit and its surrounding filed in each cavity. Therefore, the less entanglement between qubit and field (in each cavity), the less swapped entanglement between qubits. Thanks to the results presented in section \ref{secAtFiEnt}, we already know the positive role of movement of qubit on the entanglement between the qubit and the cavity field. This means that, by choosing a suitable value of $\beta$, a long-living stationary entangled state between two qubits can be created. 

\begin{figure}[h!]
\centering
\subfigure[\label{Fig6a}  $\beta=0$ ]{\includegraphics[width=0.4\textwidth]{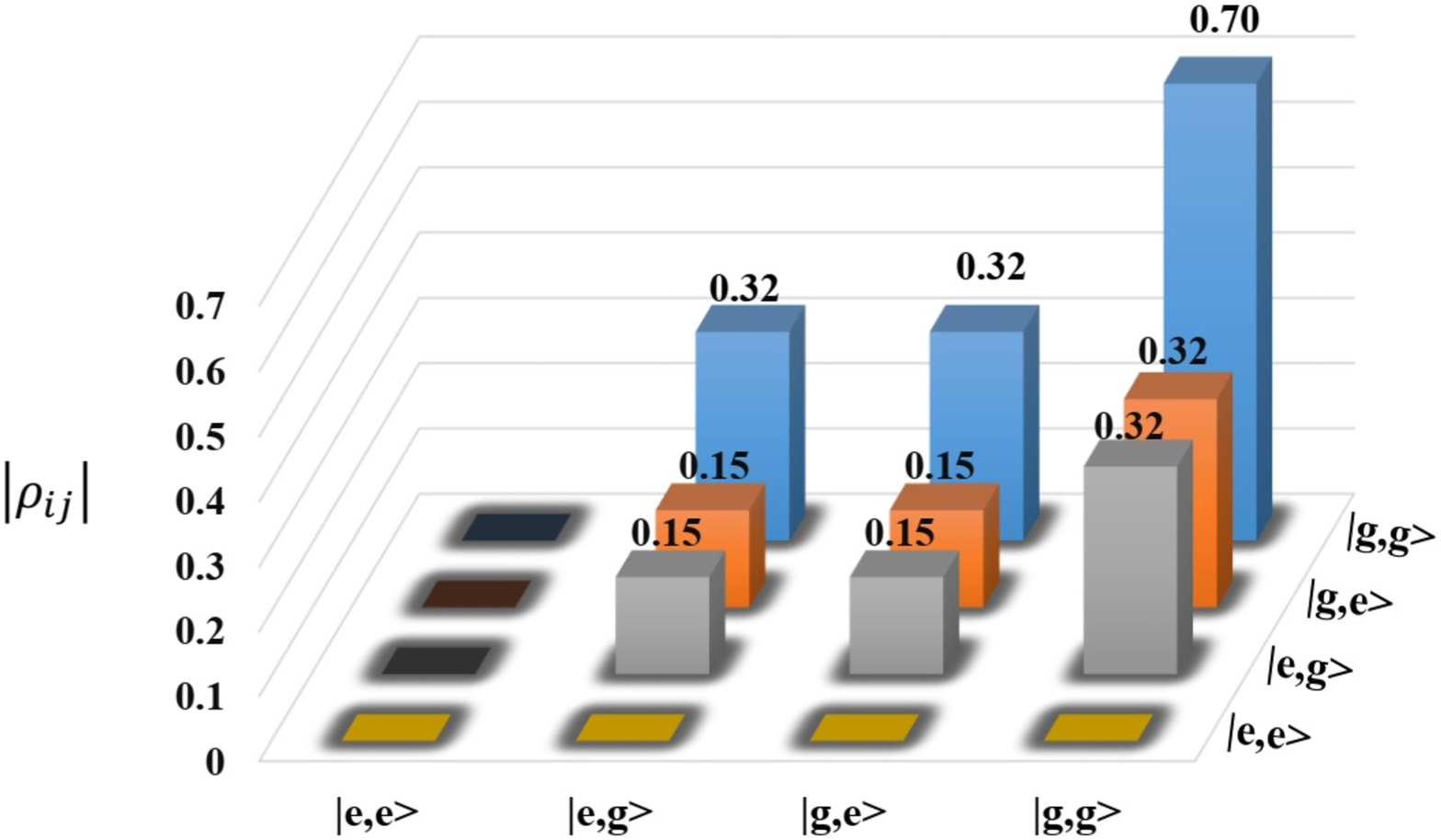}}
\hspace{0.05\textwidth}
\subfigure[\label{Fig6b} $\beta=15\times 10^{-9}$]{\includegraphics[width=0.4\textwidth]{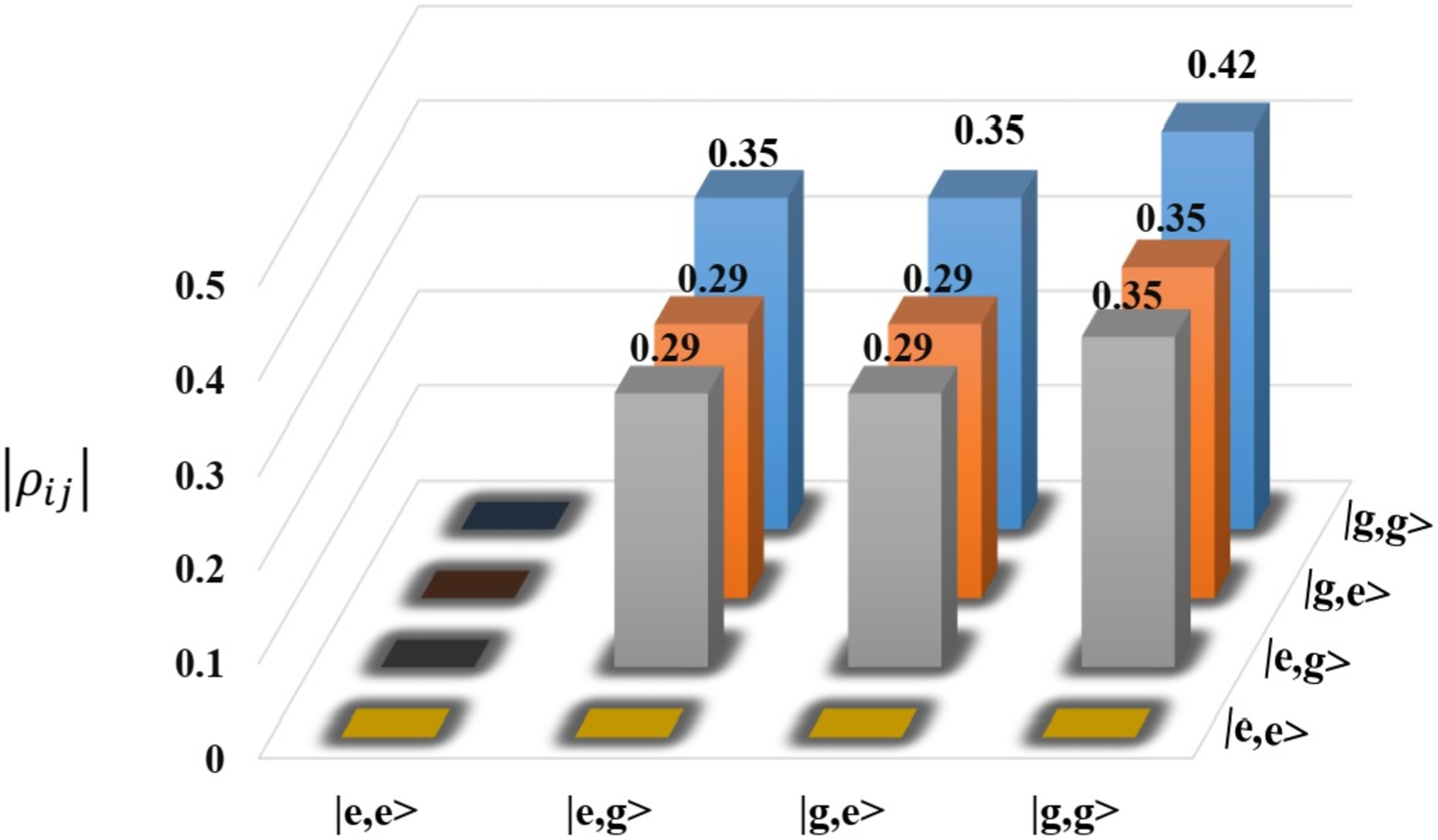}}

\caption{Histogram of density matrix of two qubits at scaled time $\tau=1$ with initial state $\ket{\Psi_\mathrm{AA}(0)} =\ket{e}\otimes\frac{1}{\sqrt{2}}\big(\ket{e}+\ket{g}\big)$ for strong coupling regime, i.e., $R=10$ with (a) $\beta=0$ and (b) $\beta=15\times 10^{-9}$. In these plots  $\omega_0/\lambda=1.5\times 10^{9} $.} \label{Fig6}
   \end{figure}

In order to explore a deep insight about the role of movement of the qubits on the swapped entanglement, we have shown the density matrix of two qubits at the scaled time $\tau=1$ for two values of $\beta$ (see Fig. \ref{Fig6}). We have considered the initial state of the qubits to be $\ket{\Psi_\mathrm{AA}(0)} =\ket{e}\otimes\frac{1}{\sqrt{2}}\big(\ket{e}+\ket{g}\big)$. As is observed, the population of the state $\ket{g,g}$ (which is zero at the beginning of the interaction) is increased due to the dissipation sources in the cavities. It is evident that by increasing the value of $\beta$, the population of this state is decreased, which leads to survival of swapped entanglement at more significant times.

\begin{figure}[h!]
   \centering
\includegraphics[width=0.4\textwidth]{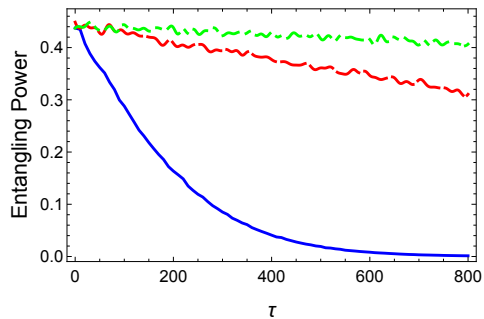}

\caption{Time evolution of the entangling power of the atom-atom state after BSM ($P^-_\mathrm{F}=\ket{\Psi^-}{}_\mathrm{FF}\bra{{}\Psi^-}$) as function of scaled time $\tau$ for weak coupling regime, i.e., $R=0.1$ with $\beta=0$ (solid black line),  $\beta=2\times 10^{-9}$ (dashed red line) and $\beta=4\times 10^{-9}$ (dot-dashed green line). We have set $\omega_0/\lambda=1.5\times 10^{9} $.} \label{Fig7}
   \end{figure}

In Fig. \ref{Fig7}, we have illustrated the entangling power  (\ref{eq:enpower}) as a function of scaled time $\tau=\lambda t$ in the weak coupling regime for various values of $\beta$. Again, the entangling power has  decaying behaviour. In this case, the oscillatory behaviour is no longer observed. The positive role of the movement of the qubits is clearly apparent.  

\begin{figure}[h]
\centering
\subfigure[\label{Fig8a}  $R=10$ ]{\includegraphics[width=0.4\textwidth]{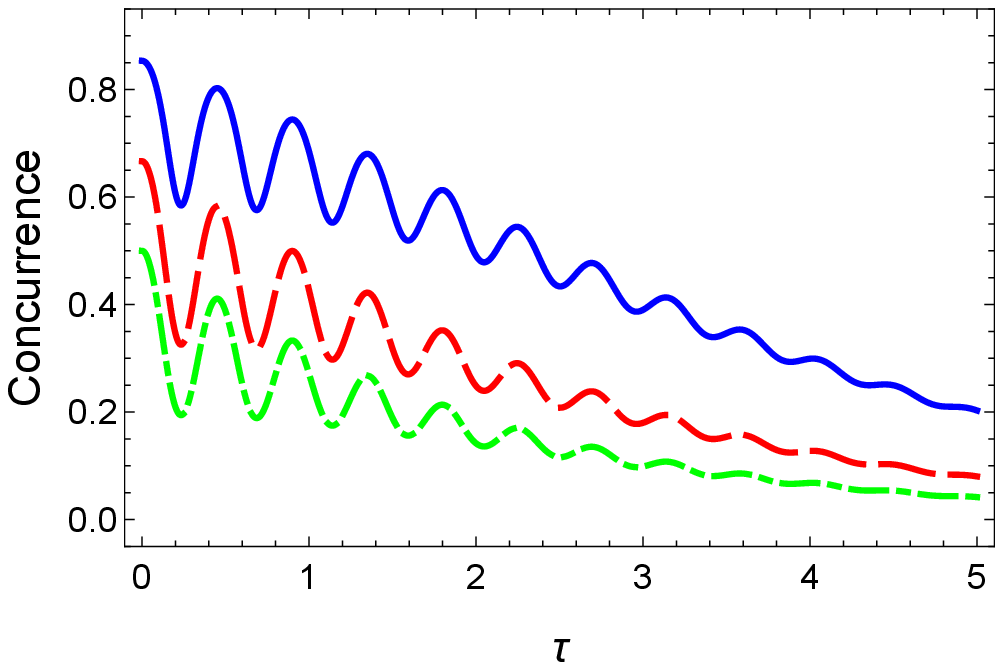}}
\hspace{0.05\textwidth}
\subfigure[\label{Fig8b} $R=0.1$]{\includegraphics[width=0.4\textwidth]{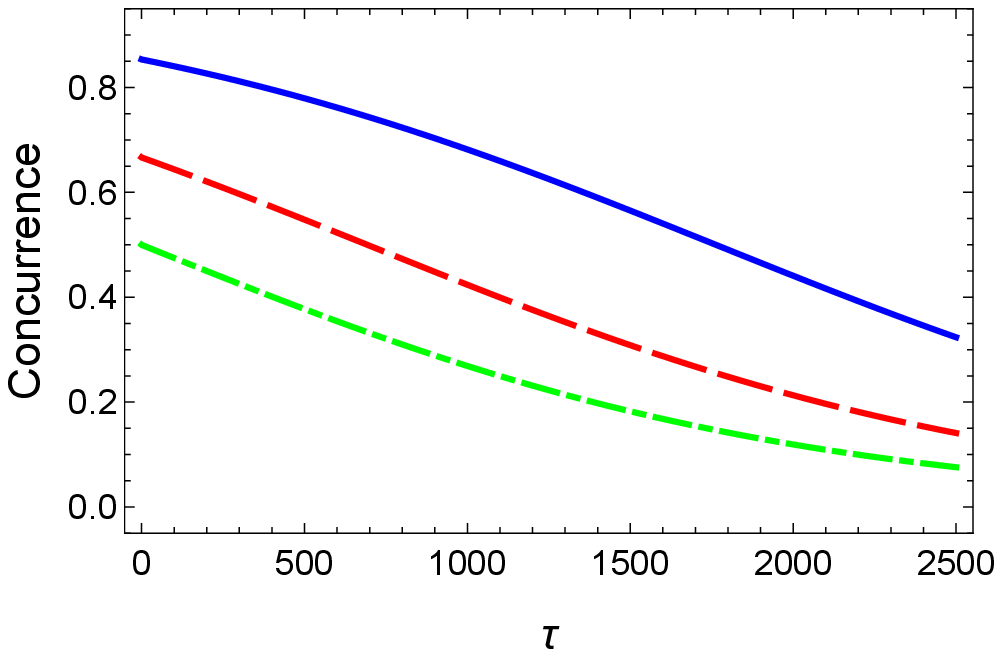}}

\caption{Time evolution of the concurrence for different initial states of the qubits for (a) $R=10$ and (b) $R=0.1$. We have set $\beta=2\times 10^{-9}$, $\omega_0/\lambda=1.5\times 10^{9} $ and $\theta_1=\frac{\pi}{2}$, $\phi_1=0$, $\theta_2=\frac{\pi}{4}$, $\phi_2=0$ (solid blue line), $\theta_1=\frac{\pi}{2}$, $\phi_1=0$, $\theta_2=0$, $\phi_2=0$ (dashed red line) and $\theta_1=\frac{\pi}{2}$, $\phi_1=\pi$, $\theta_2=\frac{\pi}{4}$, $\phi_2=0$ (dot-dashed green line).} \label{Fig8}
   \end{figure}

Finally, we take into account the effect of  initial qubit state on the evolution of the swapped entanglement. In Fig. \ref{Fig8}, we plot the time evolution of the concurrence for different initial states of the qubits. Again we consider both Markovian and non-Markovian environments. It is revealed that both coupling regimes lead to the same stationary value of entanglement. We observe  similar behaviour of entanglement for different initial states. However, the amplitude of entanglement depends on the initial state of  qubits. 

\section{Concluding Remarks}
\label{Sec.Con}

In this work, we have considered a model to study the possibility of entanglement swapping between two subsystems each contains a moving qubit inside an environment. Our model allows us to treat the environment in both weak and strong coupling regimes corresponding to bad and good cavity limits. By good cavity limit, we meant an oscillation behaviour of entanglement which is due to memory depth of  environment. We treated the movement of qubits to be classical. In the certain conditions, we have solved the time-dependent SchrÃ¶dinger equation for each subsystem. 
 
Before considering the entanglement swapping phenomenon, we have examined the influence of the movement of qubits on the entropy evolution of subsystems. This gave us an insight about possible role of the movement of qubits on the entanglement dynamics. Our parametrization for the initial state of the qubits in each subsystem allows us to construct an initial state independent for the entropy (and later on for the swapped entanglement). This has been done by taking a statistical average over all of the initial states of the qubits. The results show the entanglement between qubits and its surrounding environment due to the interaction among them. However, due to the environmental effects, the entropy has a decaying behaviour. This deterioration of entanglement is suppressed in the presence of movement of the qubit. Altogether, we recover the results presented in \cite{NourmandipourPRA2016} when we consider the qubits at rest. 

Then we turned into the problem of entanglement swapping between such two subsystem. Our goal was to swap the stored qubit-field entanglement in each subsystem into qubit-qubit and field-filed entanglements. Since the cavities are not perfect, the photons can leak out them. This allows us to perform a BSM on the photons leaving the cavities. We have obtained the analytical expression for the normalized state of qubit-qubit after BSM. We used the concurrence parameter to quantify the amount of swapped entanglement. We found several interesting and  noticeable points. First of all, by considering a suitable Bell state for field modes, the concurrence would be independent of  the shape of the incoming photons. Second, there is a set of the initial states which lead to a long-lived maximally entangled state corresponding to stationary state $\ket{\Psi^-}=\frac{1}{\sqrt{2}}(\ket{e,g}-\ket{g,e}).$ On the other hand, we have determined the initial states which lead to no entangled state at all. Therefore, the average of entanglement over all initial states of the qubits determines whether our model is a good entangler or not, the notion of entangling power. Form the information supplied, the average of swapped entanglement is always greater than zero which signifies that our proposed model is a good entangler.  The positive influence of the movement of qubits on the swapped entanglement is clearly apparent for both weak and strong coupling regimes. Since the velocities of the qubits are treated classically, therefore it is possible to reach at the velocities for qubits in which a nearly stationary swapped entanglement is obtained.  

We should emphasize that our results could be used in quantum communication applications. The idea behind them is to transmit and exchange of quantum information (entangled states) over long distances \cite{Briegel1998}. For instance, generating and swapping entanglement is at the heart of quantum repeater protocols. In this regard, since the environmental effects is always present, our results could boost the efficiency of such protocols. We also note that, our results could be useful in preparation of quantum states. For instance, in order to prepare an entangled state between two qubits which are located in their own distinct cavities, it is enough to detect a photon from a cavity. If we do not know that from which cavity the detected photon is coming, the qubits will be in an entangled state.

\end{document}